\input{aipcheck}

\documentclass[
    ,final            
  ]
  {aipproc}

\layoutstyle{6x9}

\begin{document}

\title{SPARC4\\A Simultaneous Polarimeter and Rapid Camera in 4 Bands}


\author{C.~V. Rodrigues}{address={Divis\~ao de Astrof\'\i sica - Instituto Nacional de Pesquisas Espaciais - Brazil}}
\author{F.~J. Jablonski}{address={Divis\~ao de Astrof\'\i sica - Instituto Nacional de Pesquisas Espaciais - Brazil}}
\author{K. Taylor}{address={Instruments4 - USA}}
\author{T. Dominici}{address={Laborat\'orio Nacional de Astrof\'\i sica - Brazil}}
\author{R. Laporte}{address={Divis\~ao de Astrof\'\i sica - Instituto Nacional de Pesquisas Espaciais - Brazil}}
\author{A. Pereyra}{address={Observat\'orio Nacional - Brazil}}
\author{C. Strauss}{address={Divis\~ao de Astrof\'\i sica - Instituto Nacional de Pesquisas Espaciais - Brazil}}
\author{A. M. Magalh\~aes}{address={Universidade de S\~ao Paulo - Brazil}}
\author{M. Assafin}{address={Observat\'orio do Valongo/UFRJ - Brazil}}
\author{A. Carciofi}{address={Universidade de S\~ao Paulo - Brazil}}
\author{J.~E.~R. Costa}{address={Divis\~ao de Astrof\'\i sica - Instituto Nacional de Pesquisas Espaciais - Brazil}}
\author{D. Cieslinski}{address={Divis\~ao de Astrof\'\i sica - Instituto Nacional de Pesquisas Espaciais - Brazil}}
\author{G. Franco}{address={Universidade Federal de Minas Gerais - Brazil}}
\author{A. Kanaan}{address={Universidade Federal de Santa Catarina - Brazil}}
\author{A. Milone}{address={Divis\~ao de Astrof\'\i sica - Instituto Nacional de Pesquisas Espaciais - Brazil}}
\author{K.~M.~G. Silva}{address={Divis\~ao de Astrof\'\i sica - Instituto Nacional de Pesquisas Espaciais - Brazil}}

\begin{abstract}
We present the basic concept of a new astronomical instrument: SPARC4 - Simultaneous Polarimeter and Rapid Camera in 4 bands. SPARC4 combines in one instrument: (i) photometric and polarimetric modes; (ii) sub-second time-resolution in photometric mode and excellent time-resolution in polarimetric mode; (iii) simultaneous imaging in four broad-bands for both modes. This combination will make SPARC4 a unique facility for ground-based optical observatories. Presently, the project is in its conceptual design phase. 
\end{abstract}

\maketitle


\section{Motivation to build SPARC4}

SPARC4 - \textbf{S}imultaneous \textbf{P}olarimeter \textbf{a}nd \textbf{R}apid \textbf{C}amera in \textbf{4} bands - is conceived as a new instrument for the Brazilian 1.6-m telescope of \textit{Observat\'orio do Pico dos Dias/Laborat\'orio Nacional de Astrof\'\i sica} (OPD). SPARC4 has the appropriate characteristics to improve observatory efficiency and is primarily directed towards exploring time domain variability using differential techniques to gather data simultaneously in four broad-bands. Beyond this, SPARC4 is a versatile instrument that can perform photometry, polarimetry, and astrometry. These capabilities permit an optimized use of the OPD site given its relatively modest atmospheric conditions. SPARC4 offers a new and competitive option for the suite of instruments at OPD. Data from this site still contribute for an important fraction of the Brazilian astronomical production.

\section{Science Cases}

The SPARC4 project has as its objective the supply to OPD of an efficient and versatile imager which is anticipated to address much of the demand of Brazilian astronomers in their specific areas of research. The main scientific drivers motivating SPARC4 are:

\begin{itemize}

\item the accretion structures in compact binaries;

\item the modes of pulsating stars;

\item steady and variable circumstellar envelopes, from their geometry to their composition; 

\item stellar population and star clusters;

\item magnetic field and grains in interstellar medium;

\item the origin of the variable emission from blazars;

\item transient phenomena in solar system such as transits and occultations.

\end{itemize}

\section{Instrument Overview}

\begin{figure}[bhtp]
  \includegraphics[trim = 10mm 80mm 20mm 80mm, clip,height=.40\textheight]{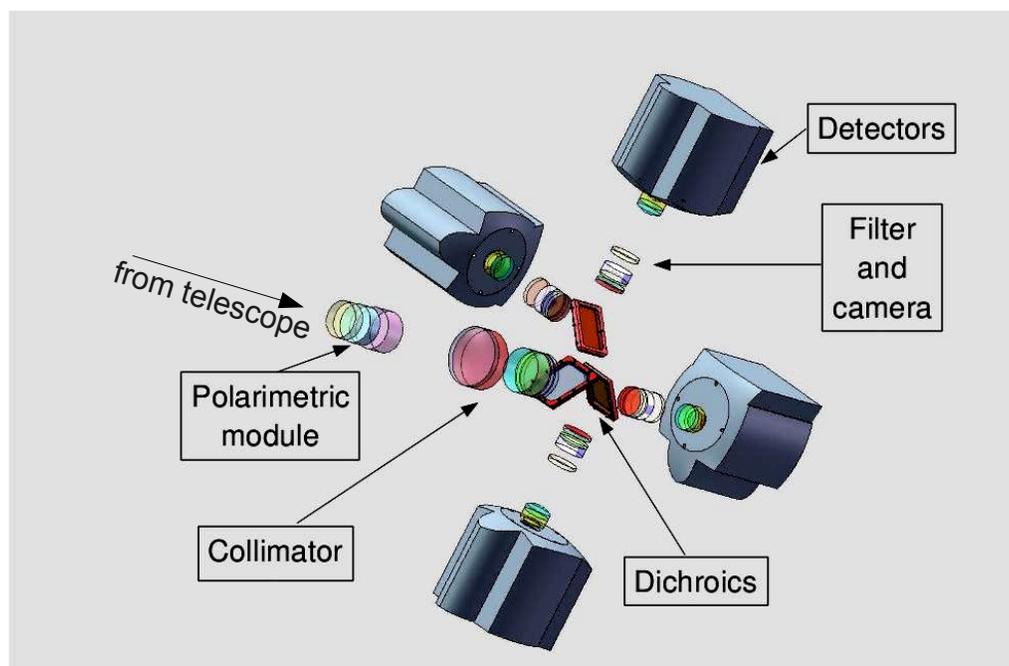}
  \caption{SPARC4 layout.}
  \label{fig_layout}
\end{figure}

SPARC4 is an imager having the following capabilities: photometric and polarimetric modes; sub-second time resolution; simultaneous imaging in 4 broad-bands in both modes. The optical path after telescope focus is, following the light: polarimetric optics, collimator, dichroics (where the beam is split in four), filters, cameras, and detectors. A preliminary layout is shown in Figure \ref{fig_layout}.

\bigskip 

The polarimetric mode is brought into play by a rotating achromatic wave-plate and an analyzer. We adopt the same configuration of the instrument described in \citet{mag1996}, which has been providing excellent results. In that design, the incoming light is separated into two beams - ordinary and extraordinary - by a Savart prism (Figure 2). The polarization of the incoming light is calculated from the relative modulation of the fluxes from the two beams. To measure linear and circular polarizations, there will be two retarder options: $\lambda$/4 or $\lambda$/2. It should be noted that the user will have the option of obtaining images without the polarimetric elements in the beam.

\begin{figure}[hbtp]
  \includegraphics[height=.15\textheight]{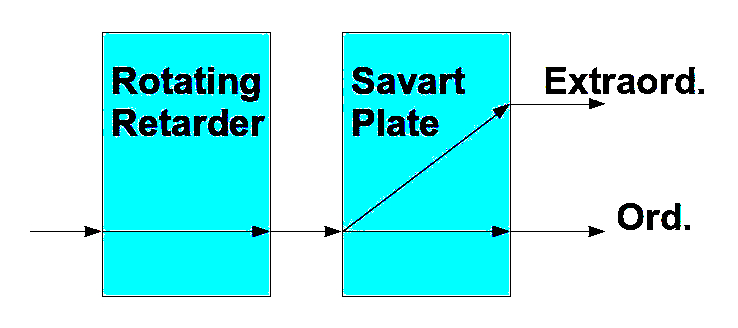}
  \caption{Schematic representation of the polarimetric module.}
  \label{fig_polar_module}
\end{figure}

After the collimator, the white light is separated in four broad-band zones closely approximating to the griz Sloan bands \citep{fu}. This is done by using three dichroics and glass filters (see Figure 1). 

At the 1.6-m OPD telescope, SPARC4's 2:1 focal reducer (f/10 to f/5) combined with a 1k detector with a pixel size of 13 micra will deliver a field of view of 5.7 arcmin square. This field of view provides an adequate number of field stars for all Galactic coordinates. This is important for differential photometry, which depends on having several comparison stars in the observed field. The scale of 0.34" per pixel is adequate for a typical seeing of 1~arcsec at the OPD site. 

The detectors should allow a time resolution of 1s or better and have a low read noise. We are studying options that include electron multiplying (EMCCD) and/or frame transfer (FT) technologies.

The instrument will include an auto-guider and will probably have an auto-focus mechanism.
 
\section{Status of the project}

The conceptual design phase has just been approved by the Brazilian funding agency Fapesp (Proc. 2010/01584-8). We expect to finish this phase in the first semester of next year. Hopefully, we will start building SPARC4 in 2012.

\begin{theacknowledgments}
SPARC4 team acknowledges the support of Fapesp (Proc. 2010/01584-8). CVR also acknowledges her grant from the Brazilian foundation CNPq (308005/2009-0).
\end{theacknowledgments}

\bibliographystyle{aipproc}   


\end{document}